\documentclass[amsmath, preprintnumbers, 10pt, twocolumn, pre bibliograpy]{revtex4-1}

\usepackage{graphicx}
\usepackage{subfigure}
\usepackage{color}
\newcommand{\MW}{\mathbf W}
\newcommand{\MA}{\mathbf A}
\newcommand{\MB}{\mathbf B}
\newcommand{\MX}{\mathbf X}
\newcommand{\MZ}{\mathbf Z}
\newcommand{\aff}{\mathcal A}
\newcommand{\R}{\mathcal R}

\renewcommand{\Re}{{\mathrm Re}}
\renewcommand{\Im}{{\mathrm Im}}
\usepackage{physics}

\begin{document}
\title{High chemical affinity increases the robustness of biochemical oscillations} 
\author{Clara \surname{del Junco} and Suriyanarayanan Vaikuntanathan}
\affiliation{Department of Chemistry and The James Franck Institute, University of Chicago, Chicago, IL, 60637}
\begin{abstract}

Biochemical oscillations are ubiquitous in nature and allow organisms to properly time their biological functions. In this paper, we consider minimal Markov state models of nonequilibrium biochemical networks that support oscillations. We obtain analytical expressions for the coherence and period of oscillations in these networks. These quantities are expected to depend on all details of the transition rates in the Markov state model. However, our analytical calculations reveal that driving the system out of equilibrium makes many of these details - specifically, the location and arrangement of the transition rates - irrelevant to the coherence and period of oscillations. This theoretical prediction is confirmed by excellent agreement with numerical results.  As a consequence, the coherence and period of oscillations can be robustly maintained in the presence of fluctuations in the irrelevant variables. While recent work has established that increasing energy consumption improves the coherence of oscillations, our findings suggest that it plays the additional role of making the coherence and the average period of oscillations robust to fluctuations in rates that can result from the noisy environment of the cell.

\end{abstract}

\maketitle

\section{Introduction}

Many organisms possess circadian rhythms, internal clocks implemented as a series of chemical reactions that result in periodic oscillations in the concentrations of certain biomolecules over the course of a day~\cite{Johnson2017, Panda2002}. These oscillations allow organisms to time their biological functions in synchrony with changes in daylight, thereby increasing the fitness of the organism~\cite{Novak2008, Woelfle2004, Johnson2017}. Yet, each chemical reaction underlying a biochemical oscillator is a stochastic process, which leads to fluctuations in the period of oscillations and affects how accurately it can tell time. In addition to this intrinsic noise, the heterogeneous environment inside a cell can increase the uncertainty in the clock's period~\cite{Pittayakanchit2018}. Understanding how biological organisms can robustly maintain the time scales of their clocks in the presence of these fluctuations is hence a central question~\cite{Barkai2000, Barato2017, Barato2015, Gingrich2016, Morelli2007}, which we address in this letter.  Our main result shows that nonequilibrium driving can dramatically reduce the number of parameters that control oscillator time scales. Oscillator time scales thus become insensitive to changes in many parameters, making them robust and tunable even when the reaction rates underlying the oscillator contain significant heterogeneity and are not irreversible.

\begin{figure}
\centering
\includegraphics[width=0.9\linewidth]{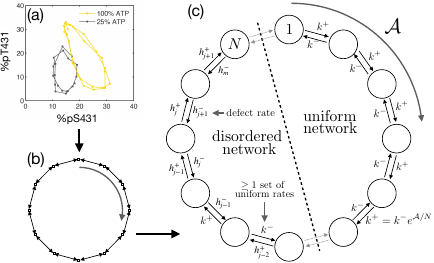}
 \caption{The model for a biochemical oscillator studied in this paper. (a) Biochemical oscillators trace a limit cycle in a high-dimensional state space of chemical concentrations. In this example of the KaiC oscillator, reprinted from Ref.~\citenum{Leypunskiy2017}, the axes represent the fraction of KaiC phosphorylated at each of two sites. (b) We approximate these limit cycles by projecting them down on to a single cycle of states. (c) Each node represents a state of the system. The system hops between states with rates $k_i^\pm$ and is driven out of equilibrium by an affinity $\aff$. This creates a net clockwise current, resulting in damped oscillations in the probability of finding the system in a particular state. We probe the dependence of the oscillation time scales on the rates of the network by adding `defect rates' $h_j^\pm$ to the uniform network, for which the time scales are known, and obtaining expressions for the coherence and period of oscillations in the disordered network. Our theory works in disordered networks where no two rates are equal.}
\label{fig:network}
\end{figure}

The model we use to derive our results (Fig.~\ref{fig:network}) is motivated by the fact that in a general sense oscillators undergo (noisy) limit cycles. The model consists of $N$ states connected in a ring that represents a projection of an oscillator's average limit cycle. For instance, in the well-studied KaiABC oscillator of the cyanobacteria {\it S. elongatus}, each of these states would represent a vector of counts of the different phosphorylation states of a population of KaiC proteins~\cite{Nakajima2005,Rust2007, VanZon2007, Marsland2019}. The system can hop between states with rates $k_i^{\pm}$, which could represent (de)phosphorylation rates. The source of oscillations is that the forward reaction rates in the KaiABC cycle are larger than the reverse rates. The rates in our model reproduce this asymmetry, creating a nonequilibrium steady state with a net clockwise current~\cite{Seifert2012}. The chemical driving force responsible for the current can be quantified by the ``affinity" of the network, $\aff \equiv \log \prod_{i=1}^N k_i^+/k_i^-$~\cite{Seifert2012}. In the case of KaiC, which is an ATPase, the affinity is provided by the highly exergonic hydrolysis of ATP~\cite{Terauchi2007}.  If the system is initialized on a state $i_{\rm 0}$ in a network with a non-zero affinity, the probability associated with finding the system in any state will exhibit damped oscillations. The period of the oscillations reflects the average time taken by the system to traverse the ring and return to the state $i_0$. The damping in the oscillations is an unavoidable consequence of the stochastic nature of the transitions. The ratio $\R$ of the damping time to the oscillation time provides a figure of merit for the coherence of oscillations in the network~\cite{Qian2000b, Cao2015, Morelli2007,Cao2015, Nguyen2018}. 

In principle, $\R$ depends on all the details of the rates $k_i^\pm$ in the network. However, in line with a large body of work that generically connects energy dissipation to accuracy in biophysical processes~\cite{Hopfield1974, Bennett1979, Qian2006, Lan2012, Mehta2012, Murugan2012}, it has been suggested that irrespective of these details the affinity bounds the coherence of biochemical oscillations~\cite{Cao2015, Barato2017, Wierenga2018, Nguyen2018}. 
In particular, Barato and Seifert recently conjectured an upper bound on $\R$ as a function of the number of states $N$ and the affinity $\mathcal A$ of the biochemical network~\cite{Barato2017}. The bound is saturated when the network is uniform; that is, when all of the counterclockwise (CCW) rates in the network are equal and all of the clockwise (CW) rates are equal.  However, the bound is a weak constraint for non-uniform networks with arbitrary rates~\cite{Barato2017, Cao2015} and hence it is unclear which variables control the time scales of the oscillator in the presence of rate fluctuations. If the time scales depend sensitively on all of the rates in the network, then they will vary dramatically with any fluctuations in the rates. Conversely, if the time scales depend on only a small subset of the variables, then they will be robust to any fluctuations that do not affect this subset.
We probe this question by obtaining analytical expressions for the time scales of Markov state models with non-uniform rates. 
Our main result, derived in section~\ref{sec:theory}, is an analytical expression for an eigenvalue of the transition rate matrices of these models, which gives the period $T$ and number of coherent oscillations $\R$.  In general these quantities depend on the magnitudes and locations of the all of the entries in the transition rate matrix, yet our result depends only on the single-site distribution of the rates. While the result is formally exact in the limit that the $\exp(-\aff/N) \to 0$, in practice our numerical results in section~\ref{sec:numerical} show that it works for surprisingly low values of the affinity.   

From a technical perspective, our results allow us to analytically compute values of the coherence and period in disordered regimes where $\R$ is significantly smaller than the upper bound.  From a biological perspective, our results suggest that as well as minimizing inherent fluctuations due to the stochasticity of the underlying processes~\cite{Cao2015, Barato2017, Barato2015, Gingrich2016}, a large energy budget has the additional, as-yet-unexplored advantage of reducing uncertainty in the presence of the additional level of disorder in reaction rates.

\section{Analytical theory for $\R$ and $T$ in disordered networks with high affinity}\label{sec:theory}

\subsection{Eigenvalues of uniform networks}\label{sec:uniform}

As in Ref.~\citenum{Barato2017}, we compute $\R$ from the ratio of the imaginary to real parts of the first non-zero eigenvalue ($\phi$) of the transition rate matrix associated with the Markov state network. We approximate the period of oscillations by $T \approx 2\pi/|\Im[\phi]|$ and the correlation time by $\tau \approx -1/\Re[\phi]$, and $\R = |\Im[\phi]|/-\Re[\phi]$. Formally, $T$ and $\tau$ depend on all the eigenvalues of the transition rate matrix, but in Appendix~\ref{sec:eigenvalue}, we show that $\phi$ captures the important features of $T$ and $\tau$. 

Ref.~\citenum{Barato2017} states that for a fixed affinity $\aff$ and number of states $N$, $\R$ is bounded by
\begin{equation}
\mathcal R \leq \cot(\pi/N)\tanh[\mathcal A/(2N)] \equiv \R_0
\label{eq:bound}
\end{equation}
and that the bound is saturated in a uniform network, that is, when $k_i^+ = k^+ = \exp(\aff/N)k^-$ and $k_i^- = k^-$ for all $i$. The transition rate matrix $\MW^{(0)}$ for the uniform network is given by 
\begin{equation}
\MW^{(0)}_{ji} = k^+\delta_{i, j-1} + k^- \delta_{i, j+1} - (k^- +k^+)\delta_{i,j}. \label{eq:w0}
\end{equation}
$\MW^{(0)}$ is a circulant matrix whose $i$th row is the top row shifted to the right by $i$ columns~\cite{Aldrovandi2001}. Its eigenvalues are the discrete Fourier transform of the first row, giving $\phi^{(0)} = -(k^+ + k^-) + k^+ e^{2\pi i /N} + k^- e^{-2\pi i /N}$ from which $\R_0$ is immediately recovered. We begin from this known result in order to find how the addition of disorder changes $\phi$. We perturb the uniform network by adding some number $m \leq N-1$ of ``defect rates" denoted $h_j^\pm$  as illustrated in Fig.~\ref{fig:network}; these could be due to some inherent asymmetry in the network (i.e. not all of the reactions making up the cycle are the same), and/or to local fluctuations in variables such as concentration that affect reaction rates. The essential insight that our derivation provides is as follows: whereas in general the new eigenvalue $\phi$, and therefore $\R$ and $T$, depend on all the details of the perturbed transition rate matrix, when the value of $\exp(-\aff/N)$ is large, significant simplifications are possible which lead to expressions that depends only on the {\it values} of the defect rates and not their {\it relative locations}.

\subsection{Transfer matrix formulation of the eigenvalue problem}\label{sec:transfer}

Rather than directly perturbing $\MW^{(0)}$, which would restrict the defect rates to be close to the uniform rates, we recast the eigenvalue problem in terms of transfer matrices~\cite{Vaikuntanathan2014a}. The transfer matrix formulation is useful for studying properties of systems with high degrees of translational symmetry and rapidly decaying spatial interactions and has been used to study localization in tight binding models~\cite{Crisanti1993} and neural networks~\cite{Amir2016}, as well as dynamic~\cite{Vaikuntanathan2014a} and structural phase transitions~\cite{Onsager1944}. 

Consider the eigenvalue equation for the circulant matrix $\MW^{(0)}$:
\begin{equation}
\MW^{(0)}   \vec f = \phi^{(0)} \vec f  \label{eq:transfer_step1}
\end{equation}
We can then write:
\begin{align}
-(k^- + k^+)f_1 + k^-f_2 + k^+f_N &= \phi^{(0)} f_1\\
 -(k^- + k^+)f_2 + k^-f_3 + k^+f_1 &= \phi^{(0)} f_2  \label{eq:transfer_step2}
\end{align}
and so forth. Solving for $f_1$ in Eq.~\ref{eq:transfer_step2} gives:
\begin{equation}
f_1 = \frac{\phi^{(0)}+ k^- + k^+}{k^+} f_2 - \frac{k^-}{k^+}f_3
\label{eq:transfer_step3}
\end{equation}
which we can also write as:
\begin{equation}
\begin{bmatrix}
f_1 \\ f_2
\end{bmatrix}
= 
\begin{bmatrix}
\frac{\phi^{(0)}+ k^- + k^+}{k^+} & - \frac{k^-}{k^+} \\
1 & 0
\end{bmatrix}
\begin{bmatrix}
f_2 \\ f_3
\end{bmatrix}
\equiv
\mathbf B_0 
\begin{bmatrix}
f_2 \\ f_3
\end{bmatrix}
\label{eq:transfer_step4}.
\end{equation}
Thus, $\mathbf B_0$ maps the eigenvector magnitudes $(f_{i-1}, f_{i})$ to $(f_{i}, f_{i+1})$. Because the matrix $\mathbf B_0$ is the same for each link in the unicyclic network with uniform rates, we have:
\begin{equation}
\begin{bmatrix}
f_1 \\ f_2
\end{bmatrix}
=
\mathbf B_0^N
\begin{bmatrix}
f_1 \\ f_2
\end{bmatrix}
\label{eq:transfer_step5}
\end{equation}
so that $\mathbf B^N$ must have an eigenvalue of 1.  Solving for the eigenvalues of $\mathbf B_0^N$ will give a polynomial of order $(\phi^{(0)})^N$, the $N$ roots of which are the $N$ eigenvalues of the transition matrix $\MW_0$. This gives us an alternative to Eq.~\ref{eq:transfer_step1} for finding $\phi^{(0)}$.

\subsection{One defect rate}\label{sec:onedefect}

We first consider how adding one set of ``defect" rates $h^\pm$ shifts $\phi$. To do so we replace one of the $\mathbf B_0$ matrices in the product in Eq.~\ref{eq:transfer_step5} by
\begin{equation}
\MA \equiv 
\begin{bmatrix}
\frac{\phi+ h^- + h^+}{h^+} & - \frac{h^-}{h^+} \\
1 & 0
\end{bmatrix}.
\end{equation}
which maps the eigenvector elements on either side of the link with the defect rates, so that now the eigenvalue of $\MA \MB^{N-1}$ is equal to 1. Now the $\mathbf B$ matrix has changed, since modifying the rates changes the value of $\phi$. We write $\phi$ in the most general way as:
\begin{equation}
\phi = \phi^{(0)} + C\gamma
\label{eq:phipert}
\end{equation}
where $C$ is a constant to be determined and $\gamma$ is an unknown to be calculated. This implies 
\begin{equation}
\MB = \MB_0 + \begin{bmatrix} C\gamma/(k^+) &  0 \\ 0 & 0 \end{bmatrix} \equiv \MB_0 + \MB_1.
\end{equation}

We now proceed with the matrix perturbation of $\mathbf B$. First we compute the eigenvalues ($\beta_i$) and normalized eigenvectors of $\mathbf B_0$. Note that since $\mathbf B_0$ is non-Hermitian, its right and left eigenvectors ($\bra{i}$ and $\ket{i}$) are not the same. We obtain:
\begin{align}
\beta_1^0&= e^{2\pi i/N} &&\beta_2^0 = (k^-/k^+)e^{-2\pi i/N} \\
\ket{1_0} &= \frac{1}{c_1}(\beta_1^0, 1) &&\ket{2_0} =\frac{1}{c_2}(\beta_2^0, 1) \label{eq:eigv1} \\
\bra{1_0} &= \frac{1}{c_1}(-1/\beta_2^0, 1) && \bra{2_0} = \frac{1}{c_2}(-1/\beta_1^0,1) \label{eq:eigv2} \\
c_1^2 & = 1-\beta_1^0/\beta_2^0  && c_2^2 = 1-\beta_2^0/\beta_1^0.
\end{align}
Now we compute the first-order correction to the eigenvalues:
\begin{equation}
\beta_1^{(1)} = \mel{1_0}{\mathbf B_1}{1_0} =-\frac{e^{4\pi i /N}}{c_1^2}\frac{C\gamma}{k^-}.
\label{eq:PT1}
\end{equation}
We choose
\begin{equation}
C = -c_1^2 k^- e^{-2\pi i/N}
\label{eq:constant}
\end{equation}
so that
  \begin{equation}
 \beta_1 = e^{2\pi i/N} (1 + \gamma).
 \end{equation}
 Similarly,
  \begin{equation}
 \beta_2 = \frac{k^-}{k^+}e^{-2\pi i /N}(1 + \gamma) =  e^{-\aff/N}e^{-2\pi i /N}(1 + \gamma).
 \end{equation}
 We show in Appendix~\ref{sec:exacteig} that these expressions are exact in the high affinity limit where $\exp(-\aff/N) \to 0$.
We can now compute $\mathbf B^{N-1}$ using:
 \begin{equation}
\mathbf B^{N-1} =\sum_i \beta_i^{N-1}\MX^{0}_i
 \end{equation}
where $\MX^{0}_i\equiv \ket{i_0}\bra{i_0}$. 

Since $\beta_2^{(1)} \propto e^{-\aff/N} < 1$,  if $\aff/N$  is sufficiently large, all of the terms containing $\beta_2$ will vanish. Then the product $\MA \MB^{N-1}$ becomes:
\begin{equation}
\MA \MB^{N-1} = \beta_1^{N-1}\MA \MX_1^{(0)}  \equiv \beta_1^{N-1}\MZ
\label{eq:ABreduced}
\end{equation}

We now compute this product and set its eigenvalue equal to 1 in order to solve for $\phi$. 
We find:
\begin{equation}
\mathbf Z  = \frac{1}{c_1^2}
\begin{bmatrix}
d a & d b \\ a & b
\end{bmatrix}
\label{eq:z}
\end{equation}
where 
\begin{align}
d(h^\pm)  &= \frac{ \gamma+e^{-2 i \pi /N} (k^--h^-) +k^+ e^{2 i \pi/N}+h^-+h^+-k^--k^+}{h^+} \\
a &= -e^{4\pi i/N}k^+/k^- \\
b & = e^{2\pi i /N}
\end{align}

Since the two rows of $\mathbf Z$ are related by a constant, $\mathbf Z$ has a zero eigenvalue. The non-trivial eigenvalue $\zeta$ of $\mathbf Z$ is:
\begin{widetext}
\begin{equation}
\zeta = \frac{e^{2\pi i/N} \left({k^+} e^{2\pi i/N} (-C\gamma-{h^-}-{h^+}+{k^-}+{k^+})+{h^-} {k^+}+{h^+} {k^-}-{k^-} {k^+} -e^{4\pi i/N}{k^+}^2\right)}{c_1^2 {h^+} {k^-}}.
\label{eq:zeta}
\end{equation}
\end{widetext}
We can now solve for $\gamma$ using:
\begin{align}
1 & =\beta_1^{N-1}  \zeta \\
& = e^{-2\pi i/N} (1 +\gamma)^{N-1} \zeta
\end{align}
For notational simplicity we absorb the $e^{-2\pi i/N}$ term in to $\zeta$, letting 
\begin{equation}
\zeta' = e^{-2\pi i/N}\zeta.
\label{eq:zetaprime}
\end{equation}
Rearranging, we have:
\begin{equation}
(1 + \gamma)=  \zeta'^{1/(1-N)}
\end{equation}
We now rewrite $\zeta'^{1/(1-N)}$ as $\exp(\log(\zeta'^{1/(1-N)})) = \exp(\left(\frac{1}{1-N}\right)\log(\zeta'))$ and expand to second order, giving
\begin{equation}
\gamma \approx  \frac{1}{1-N}\log(\zeta')  + \frac{1}{2(1-N)^2}(\log(\zeta'))^2. \label{eq:gamma-self-cons}
\end{equation}

This gives a self-consistent equation for $\gamma$ (since $\zeta'$ is a function of $\gamma$). To obtain the numerical results presented later in this paper, we solved Eq.~\ref{eq:gamma-self-cons} numerically by searching for roots of the equation in the neighborhood of an explicit approximation for $\gamma$ that we obtain by expanding the logarithm in Eq.~\ref{eq:gamma-self-cons} to linear order (see Appendix~\ref{sec:linexp}). 

\subsection{Many Defect Rates}\label{sec:manydefects}

We can extend Eq.~\label{eq:gamma-self-cons} to the case where where many ($m$) of the rates are `defects'.  The product of transfer matrices in this case is:
\begin{equation}
\begin{bmatrix}
f_1 \\ f_2
\end{bmatrix}
=
\MA_{1}\MB^{L_1}\cdots \MA_{j}\MB^{L_j} \cdots \MA_{m}\MB^{L_m}
\begin{bmatrix}
f_1 \\ f_2
\end{bmatrix}.
\label{eq:ABmult}
\end{equation}
where $L_j$ is the distance (the number of uniform rates) between neighboring defect rates.  Generally, Eq.~\ref{eq:ABmult} has $2^m$ terms. However, if $e^{-L_j\aff/N}$ is sufficiently large, we can ignore $\beta_2$ as we did in the case of one defect above. Then, Eq.~\ref{eq:ABmult} reduces to a single term from which we can factorize $\beta_1$, giving:
\begin{equation}
\begin{bmatrix}
f_1 \\ f_2
\end{bmatrix}
=
\beta_1^{N-m}\MA_{1}\MX_1^{(0)}\cdots \MA_{j}\MX_1^{(0)} \cdots \MA_{m}\MX_1^{(0)}
\begin{bmatrix}
f_1 \\ f_2
\end{bmatrix}
\label{eq:ABmultreduced}
\end{equation}
The affinity thus sets a correlation length for the defects; if the affinity per site ($\aff/N$) is sufficiently large, the spacing between them does not matter. In principle, the order of the matrices in the matrix product in Eq.~\ref{eq:ABmultreduced} is still important and hence the value of $\phi$ depends on the order of the defects. However, our calculations are simplified due to the special symmetry of $\mathbf Z_j \equiv \MA_j\MX_1^{(0)}$ (given by Eq.~\ref{eq:z} with the defect rates $h^\pm$ now indexed $h_j^\pm$, etc.). We find that the non-trivial eigenvalue of the product $\mathbf Z_i \mathbf Z_j$ is the product of the non-trivial eigenvalues of $\mathbf Z_i$ and $\mathbf Z_j$.  As a result, the expression for $\phi$ is simply determined by the product of the non-trivial eigenvalues of the $\mathbf Z_j$ matrices. Therefore, as long as $L_j \geq 1 \forall j$, the order in which the defects are placed and the spacing between them becomes irrelevant as far as $\phi$ is concerned.  We can thus simply extend our results for one defect to write:
\begin{align}
\phi &= \phi^{(0)} + \gamma \label{eq:phinew} \\
\gamma &\approx  \frac{1}{m-N}\sum_{j=1}^m \log(\zeta_j') + \frac{1}{2(m-N)^2}\left(\sum_{j=1}^m \log(\zeta_j')\right)^2
\label{eq:gamma}
\end{align}
where $\zeta'_j$ is a function of $k^+, k^-, h^+_j, h^-_j, N$ given by Eq.~\ref{eq:zeta} and~\ref{eq:zetaprime} but is independent of any of the other defect rates.


This derivation requires $L_j\geq 1$ because the non-trivial eigenvalue of the product $\mathbf Z_i \mathbf Z_j$, where $\mathbf Z_j \equiv \MA_j\MX_1^{(0)}$, is the product of the non-trivial eigenvalues of $\mathbf Z_i$ and $\mathbf Z_j$.  However, the eigenvalue of the product of $\MA_i\MA_j$ is $not$ the product of their eigenvalues. Therefore, Eq.~\ref{eq:gamma} should not be valid if there are defect rates on either side of the same node in the network ($L_j = 0$). Nonetheless, our numerical results in the main text show that these expressions accurately predict the eigenvalues of the oscillator even when $L_j = 0$ for nearly all of the defects. In Appendix~\ref{sec:lj0}, we show how high affinity makes this possible.

While Eqs.~\ref{eq:phinew} and~\ref{eq:gamma} are only formally correct in the limit of high affinity, in practice, as we show in Figs.~\ref{fig:rtvsm} and~\ref{fig:rtvssigma}, they predict $\phi$ well even for rather small values of $\aff$. Specifically, $N/\aff$ sets a length scale for correlations between defect rates, so that if $N/\aff \ll 1$ the relative positions of defect rates do not affect $\phi$. Generally, $\phi$ depends on $\mathcal O (N)$ parameters, including the values and locations of the uniform and defect rates in the model.  Eqs.~\ref{eq:phinew} - \ref{eq:gamma} predict that in the limit $N/\aff \ll 1$, $\phi$ depends only on the values of the rates. 
As such, changes in the remaining, irrelevant parameters will not affect oscillator timescales - in other words, the oscillator is robust to these changes.
As we demonstrate below, this implies that the period and coherence of oscillations can be maintained even in the presence of disorder in the rates. 

\section{Numerical results}\label{sec:numerical}

\subsection{Accurate theoretical predictions with strong nonequilibrium driving}
To test the limits of Eq.~\ref{eq:phinew}, we compared it to the result of numerical diagonalization for networks of size $N = 100$ with up to 99 defect rates placed at random locations in the network. We considered networks with quenched disorder: we set all CCW rates to $k^- = h_j^- = 1$ and randomly selected the CW defect rates $h_j^+$ from a Gaussian probability distribution $P_G(\tilde\sigma,\aff_0, N)$ with mean $k^+ = \exp(\aff_0/N)$ and standard deviation $\sigma = \tilde \sigma \exp(\aff_0/N)$, and with a lower cutoff at 0.1 so that we do not select rates that are very close to zero or negative. This prescription naturally allows the affinity to vary between networks - we show results for networks with fixed affinity in Fig.~\ref{fig:rtvsm-constaff} (in the appendix) that confirm the bound in Eq.~\ref{eq:bound}. Fig.~\ref{fig:rtvsm} shows the importance of a high affinity ($\aff_0/N$, the value of the affinity in the uniform network) for controlling $\R$ and $T$.  Our prediction from Eq.~\ref{eq:phinew} improves with increasing $\aff_0/N$: for $\aff_0/N = 2$, Fig.~\ref{fig:rtvsm} shows that Eq.~\ref{eq:gamma} is accurate even when all but one of the rates in the network are random and $\langle\R\rangle$ is 40\% less than the bound $\R_0$. 
(For comparison, the cycle of the KaiC hexamer has $\aff/N \gtrsim 10$~\cite{Rust2007, Terauchi2007, cellbionumbers}.) 

 While minimizing phase diffusion and thereby maximizing $\R$ is a priority for a biochemical clock to keep time accurately, it is additionally important that $T$, the period of oscillations, be robust and tunable, for example in order to match with an external signal~\cite{Woelfle2004}. Our theory (Eq.~\ref{eq:gamma}) shows that $T$ can be reliably controlled in the high affinity limit even in the presence of substantial disorder, and we also find that the spread of $T$ values decreases significantly with increasing affinity. 

\subsection{Biologically relevant rate fluctuations} 
\begin{figure}
\centering
\includegraphics[width = \linewidth]{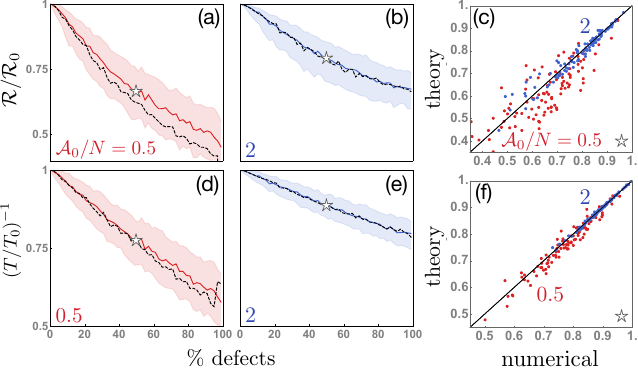}
\caption{Coherence $\R$ (a - c) and period $T$ (d - f) of an oscillator as a function of the percent of defect rates. The values of $\R$ and $T$ are more robust (spread of values decreases) and predictable at the higher affinity $\mathcal A_0/N = 2$ (b, e) than at  $\mathcal A_0/N = 0.5$ (a, d). Results are for networks with $N = 100$ states. All CCW rates are set to 1. CW defect rates $h_j^+$ are drawn from a Gaussian distribution with mean $k^+ =  \exp(\aff_0/N)$ and standard deviation $0.4k^+$. 
Because the distributions of $\R$ and $T$ are asymmetric, we plot the median (solid line) $\pm$ one quartile (shaded region) of the numerical values for 500 samples of defect rates. The dashed lines are the median theoretical predictions for 500 samples of defect rates. For $\aff_0/N = 2$ (blue), our theory is accurate even when \% defects $\approx 100$. On the right, we compare the values of (c) $\R/\R_0$ and (f) $T_0/T$ for individual realizations of the random networks with 50\% defects (parameters indicated by a star).}
\label{fig:rtvsm}
\end{figure}

\begin{figure}
\centering
\includegraphics[width = \linewidth]{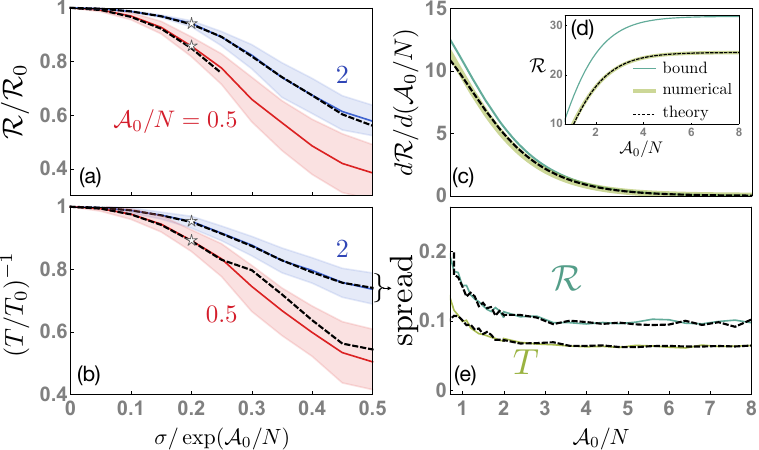}
\caption{(a) $\R$ and (b) $T$ for totally disordered networks of size $N = 100$ as a function of the standard deviation of the distribution of defect rates. All CCW rates are set to 1. CW defect rates $h_j^+$ are drawn from a Gaussian distribution with mean $k^+ =  \exp(\aff_0/N)$ and standard deviation $\sigma = \tilde\sigma k^+$.  Black dashed lines are theoretical predictions. (c - d) The absolute value of $\R$ as a function of $\aff_0/N$ plateaus in a totally disordered network (here we show results for $\tilde\sigma = 0.3$) as well as in uniform networks (given by the bound in Eq.~\ref{eq:bound}). As a result, $d\R/d(\aff_0/N)$ goes smoothly to zero in the disordered and uniform networks, even though $\R$ is far from the bound. 
(e) The `spread', defined as the distance between the median $\pm$ 1 quartile of the data, as a function of the affinity. As predicted, it decreases with increasing affinity.}
\label{fig:rtvssigma}
\end{figure}

\begin{figure}
\centering
\includegraphics[width = \linewidth]{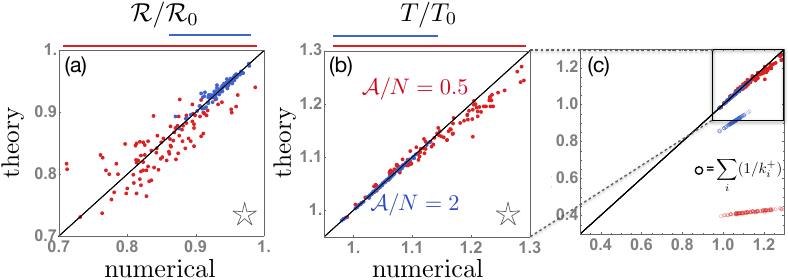}
\caption{Comparing the values of (a) $\R/\R_0$ and (b) $T/T_0$ for individual realizations of random networks. Each point represents the period of a network with randomly generated rates with parameters indicated by a star in Fig.~\ref{fig:rtvssigma}.  Open circles in the zoomed-out $T/T_0$ scatter plot (c) represent the approximation in the limit where reverse steps can be neglected ($T = \sum_i 1/k_i^+$). At $\mathcal A_0/N = 0.5$, where neglecting reverse steps underestimates the period by more than 50\%, our theory provides an accurate estimate. Note that in Fig.~\ref{fig:rtvssigma} we plotted $(T/T_0)^{-1}$; here we plot $(T/T_0)$ as it results in a more compact plot.}
\label{fig:scatterplots}
\end{figure}

The ``number of defect rates" is a convenient measure of disorder to use in the context of our theory, but is not clearly related to a biological scenario. In general we would expect all rates to be different from one another, even in the case where the underlying reaction network is relatively simple, as in the KaiC oscillator illustrated in Fig.~\ref{fig:network}. This is because the rates of the oscillator depend on the collective state of the system which is always changing~\cite{Marsland2019}, and in addition can be affected by local fluctuations, for instance in the concentration of ATP. In Fig.~\ref{fig:rtvsm} we showed that our perturbation theory can handle networks where no two rates are the same. In Fig.~\ref{fig:rtvssigma}, we further investigate these fully disordered networks by showing how $\R$ and $T$ vary as a function of the spread of rates $\sigma$ in a network with all CCW rates set to 1 and all CW rates drawn from the distribution $P_G(\tilde\sigma,\aff_0, N)$ (defined in the previous section). All of our findings still hold: the prediction becomes more accurate (Fig.~\ref{fig:rtvssigma}a, b) and the spread of $T$ and $\R$ values decrease (Fig.~\ref{fig:rtvssigma}d) as the affinity $\aff_0$ increases. In these networks the affinity naturally varies, and the average time scales are robust to these small variations. In Fig.~\ref{fig:scatterplots}, we show scatter plots comparing numerical values of the period with our analytical guess in specific realizations of the network. This data shows that we are not only able to predict the median values shown in Fig.~\ref{fig:rtvssigma}. Rather, our accurate prediction of the statistics of $\R$ and $T$ for an ensemble of oscillators is due to the success of our theory at accurately predicting the timescales for individual oscillators.

Finally, we consider changes in the uniform rate, or $\aff_0/N$. The bound in Eq.~\ref{eq:bound} implies that in a uniform network, the coherence $\R$ becomes insensitive to changes in the affinity at high $\aff_0/N$ (Fig.~\ref{fig:rtvssigma}c). However, it is not clear whether this will be the case in a disordered network. In Fig.~\ref{fig:rtvssigma}, we show that even in a disordered network where $\R < \R_0$, the dependence of $\R$ on $\aff_0/N$ vanishes smoothly for values of $\aff_0/N$ greater than  $\sim 5$. In this regime, our analytical results demonstrate how - due to nonequilibrium driving - the coherence is insensitive to large global fluctuations in the affinity that change the average rate $k^+$ as well as to small local fluctuations that cause the rates to fluctuate about $k^+$.

\section{Discussion and Conclusions} 
Biochemical oscillators, which can function as internal clocks, operate in noisy environments that can affect the ability of the clock to tell time accurately; yet somehow these oscillations continue with a well-defined period over long times. Here, we present analytical calculations supported by numerical results that show how a biochemical oscillator modeled as a Markov jump process on a ring of states (Fig.~\ref{fig:network}) can use high chemical affinity (for instance, in the form of ATP) to robustly maintain and tune its time scales even in the presence of a substantial amount of disorder. While previous work has postulated an upper bound on the number of coherent oscillations that such a model can support in terms of the chemical affinity, the bound can be loose, and does not elucidate the dependence on the details of the rates in the network~\cite{Barato2017}.  We close this gap by showing how affinity can dramatically decrease the number of relevant variables controlling the coherence and period of oscillations. 

Specifically, we consider Markov state networks such as those in Fig.~\ref{fig:network}c and sample the rates from a probability distribution in order to mimic disorder in biological systems. We consider networks with quenched disorder in order to probe the synchronization between multiple noninteracting oscillators (e.g. in different cells~\cite{Mihalcescu2004}), and as a proxy for the variation in a single oscillator's period when the rates are fluctuating over time.  Our analytical theory in Eqs.~\ref{eq:phinew} - \ref{eq:gamma} reveals that in the limit of high affinity the arrangement of the rates becomes irrelevant, and the period of oscillations $T$  and the number of coherent oscillations $\R$ depend only on the magnitude of the rates. As a result, in the limit of high affinity, we can accurately predict $\R$ and $T$ for individual realizations of the rate disorder knowing only the values of the rates. We note that while our analytical results are formally valid only when $\exp(-\aff/N) \to 0$, they do not completely neglect the possibility of reverse transitions on the network, where the period of the oscillator is simply $T = \sum_i 1/k_i^+$. In Fig.~\ref{fig:scatterplots} we compare our theory to this trivial approximation and show that our prediction is quite accurate at small values of $\aff/N$ where the trivial approximation underestimates the period by over 50\%. We also wish to contrast our results with expressions for the velocity and diffusion coefficients for periodic, disordered, one-dimensional Markov networks that have been derived previously in the literature~\cite{Derrida1983}. First, in this paper we formally calculate a different quantity: an eigenvalue, as opposed to a first passage time.  Second, these expressions in Ref.~\cite{Derrida1983} include positional correlations between rates, which can modify the transport properties; simplified expressions require an assumption that the rates are uncorrelated.  Our work shows how non-equilibrium driving can make these correlations irrelevant.

A consequence of our prediction is that for a given probability distribution of the rates, the possible values of $\R$ and $T$ for different realizations of the rates are more narrowly distributed about their average values because different arrangements of the rates no longer affect the time scales. Taken together, this means that when the affinity is high $\R$ and $T$ can be well-approximated from a small, fixed number of parameters that does not scale with system size - namely, enough to specify the single-site probability distribution of the rates - as opposed to $\mathcal O(N)$ parameters required to specify the values and locations of all the rates that are generally expected to determine $\R$ and $T$. Our results give insight in to why biochemical oscillators might evolve to consume large amounts of energy in the form of ATP~\cite{Terauchi2007}: in addition to the previously known function of suppressing uncertainty in the period of the oscillator for a system with uniform rates~\cite{Cao2015, Barato2017}, it also makes the time scales of the oscillator more robust to fluctuations in the rates caused by the noisy environment of the cell by decoupling these rates.

Although we used the example of KaiABC in this work, the Markov cycle representation is not limited to post-translational oscillators with conserved protein copy numbers and the more common transcription-translation oscillators can also be studied in this paradigm. On the other hand, the generality of our results is limited by restricting ourselves to a single cycle of states, representing the average path of a limit-cycle oscillator in a high-dimensional state space. 
If the oscillator has a dominant cycle and multiple secondary cycles, we show elsewhere~\cite{DelJunco2019b} that our results can be applied by coarse-graining the secondary cycles to obtain effective rates and then using our expression in Eq.~\ref{eq:phinew} for the main cycle. 
Our theory is thus applicable to a class of networks with disordered rates and potentially many cycles which we claim are simplified but relevant caricatures of biochemical oscillators. The implications for the robustness of oscillator timescales are therefore not limited only to this single-cycle model but may be a general feature of more detailed models of biochemical oscillators.

\begin{acknowledgments}

We thank Robert Jack and Kabir Husain for helpful comments on an earlier version of this manuscript. CdJ acknowledges the support of the Natural Sciences and Engineering Research Council of Canada (NSERC). CdJ a \'et\'e financ\'ee par le Conseil de recherches en sciences naturelles et en g\'enie du Canada (CRSNG). This work was partially supported by the University of Chicago Materials Research Science and Engineering Center (MRSEC), which is funded by the National Science Foundation under award number DMR-1420709.  SV also acknowledges support from the Sloan Fellowship and the University of Chicago.

\end{acknowledgments}


%

\begin{appendix}

\section{The largest non-zero eigenvalue captures timescales of disordered oscillators}\label{sec:eigenvalue}

In a system such as the one pictured in Fig.~\ref{fig:network}, we can define the correlation function $C_{11}(t)$ as the conditional probability of the system being in state 1 at time $t$ given that it began in state 1 at time 0. It is given by the solution of the master equation:
 \begin{align}
 C_{11}(t) & \equiv [\exp(\MW t)\mathbf P (0)]_1 \\
 &= \sum_{j=0}^{N-1} P_j^{ss} e^{\phi_j t}  \label{eq:c11sum} \\
 & =  \sum_{j=0}^{N-1} P_j^{ss} e^{- \Re[\phi_j] t} (\cos[\Im[\phi_j] t] + i\sin[\Im[\phi_j] t])
 \label{eq:c11}
 \end{align}
where ${\phi_j}$ are the $N$ eigenvalues of the $N \times N$ matrix $\MW$, $P_j^{ss}$ is the steady-state probability of finding the system in state $j$, and $[...]_1$ is the first element of the vector.

To see why the first non-zero eigenvalue $\phi_1$, which we simply denote $\phi$ in the main text, is sufficient to approximate the number of coherent oscillations, we begin with the uniform case. In that case, Eq.~\ref{eq:c11sum} simplifies to:
\begin{equation}
 C_{11}(t) =  \frac{1}{N} \sum_{j=0}^{N-1}e^{\phi_j t} 
\end{equation}

The transition rate matrix $\MW_0$ for this system is a circulant matrix whose eigenvalues lie in an ellipse in the complex plane with semi-major axis $a = k^+ + k^-$ and semi-minor axis $b=k^+ - k^-$ centered on the point $(-a, 0)$. When the affinity is large and $k^+/k^- \gg 1$, this effectively becomes a circle of radius $r = k^+$ centered at $(-r, 0)$.

 The first eigenvalue is $\phi_0$ = 0, so the first term of the sum in Eq.~\ref{eq:c11sum} gives a constant contribution of $1/N$. The angle from the real axis to the $j$th eigenvalue $\phi_j$ is $2\pi j/N$. The imaginary part of $\phi_j$ is given by $r\sin(2\pi j/N)$, and the period of oscillations of the $m$th term in Eq.~\ref{eq:c11} is $T_j = 2\pi/(r\sin(2\pi j/N))$. The ratio of $T_1$ from the first non-zero eigenvalue to $T_j$ from any subsequent eigenvalue is:
\begin{equation}
\frac{T_j}{T_1} = \frac{\sin(2\pi/N)}{\sin(2\pi j/N)} \approx \frac{2\pi/N}{2\pi j/N} = \frac{1}{j}
\end{equation}
for $N \gg j$. 
The total period of the oscillations is therefore always $T_1$. Since $\Re[\phi_1] < \Re[\phi_j]$ for all $j>1$, the number of oscillations of the correlation function is given exactly by $|\Im[\phi_1]| /(-2\pi \Re[\phi_1]) =  \R/2\pi$. 

\begin{figure}[h!]
\centering
\includegraphics[width = \linewidth]{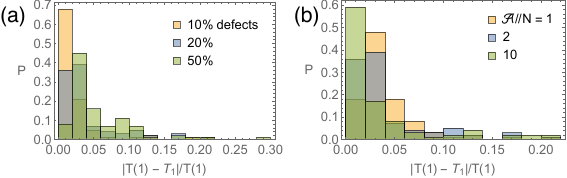}
\caption{Histograms of the relative difference between the first period of oscillations of the correlation function $C_{11}(t)$ ($T(1)$) and the period of oscillations due to the first eigenvalue  $T_1 = 2\pi/\Im[\phi]$. $T(1)$ is the location of the first peak of $C_{11}(t)$ obtained by exponentiating the full transition rate matrix. $T_1$ was calculated from the first eigenvalue, which was also obtained numerically. The data for each set of parameters were obtained from 100 randomly generated matrices (a) At constant $\aff/N = 2$ The agreement between $T(1)$ and $T_1$ gets worse with increasing disorder. (b) With the percent of defects held constant at 20\%, the agreement between $T(1)$ and $T_1$ improves with increasing $\aff/N$.}
\label{fig:periodhistograms}
\end{figure}

When defects are added the eigenvalues will no longer lie on a perfect circle in the plane, and the arguments above will no longer hold exactly. $T_j/T_1$ may no longer be an integer, so that the total period of oscillations $T\neq T_1$, and moreover the period of the oscillations at short times ($T(1)$) and the period of oscillations at long times ($T(\tau)$) may not be the same. In Fig.~\ref{fig:periodhistograms} we show histograms of the relative difference $(T(1) - T_1)/T(1)$ for different realization of matrices of size $N = 100$ with reverse rates all equal to 1, random forward rates $h_i^+$  chosen from a Gaussian distribution with mean $\mu = \exp(\aff/N)$ and variance $\sigma^2 = 0.25 \exp(\aff/N) $, and uniform forward rates $k^+$ set to maintain a constant $\aff$. We emphasize that here we are considering the difference between the first term of Eq.~\ref{eq:c11sum} and the full correlation function, both of which are obtained by numerical diagonalization; our theory presented in the main text is another level of approximation of $T_1$ on top of this.

\section{Calculation details}\label{sec:calcdetails}

\subsection{Comparing to exact eigenvalues of $\MB$}\label{sec:exacteig}

Since $\MB$ is a two by two matrix, we can compute its exact eigenvalues and check how the error in our perturbative approximations of $\beta_1$ and $\beta_2$ scales. 
The exact eigenvalues of $\MB$ are:
\begin{widetext}
\begin{align}
\beta_{\mathrm{exact}}^\pm 
 = \frac{e^{-2\pi i/N}}{2}\left(e^{-\aff/N}(1 - \gamma) + e^{4\pi i /N}(1 + \gamma) \pm \left[-4e^{4\pi i/N}e^{-\aff/N} + \left( e^{-\aff/N}(1 - \gamma) + e^{4\pi i/N}(1 + \gamma) \right)^2 \right]^{1/2}\right)
\label{betaexact}
\end{align}
\end{widetext}
If we ignore terms of order $e^{-\aff/N}$ compared to terms of order 1, the expressions above simplify to:
\begin{equation}
\lim_{\exp(-\aff/N) \to 0} \beta_{\mathrm{exact}}^\pm =  \frac{e^{2\pi i/N}}{2} \left( (1 + \gamma) \pm (1 + \gamma) \right)
\end{equation}
giving
\begin{align}
\lim_{\exp(-\aff/N) \to 0} \beta_{\mathrm{exact}}^+ = \beta_1, && \lim_{\exp(-\aff/N) \to 0} \beta_{\mathrm{exact}}^-= 0
\end{align}
In the limit of very high affinity, Eq.~\ref{eq:ABreduced} is exact. 

\subsection{Linear expansion of Eq.~\ref{eq:gamma-self-cons}}\label{sec:linexp}

To obtain an explicit expression for $\gamma$ that we use to search for roots of Eq.~\ref{eq:gamma-self-cons}, we rewrite $\zeta'$ (defined in Eq.~\ref{eq:zetaprime}) as:
\begin{equation}
\zeta' = \zeta'_0 - \gamma \frac{ {k^+} e^{2\pi i/N} C}{c_1^2 {h^+} {k^-}} = \zeta'_0 - \gamma \frac{ k^+}{ h^+ }
\end{equation}
where $ \zeta'_0 $ is independent of $\gamma$.
We then expand the logarithm as:
\begin{align}
\log(\zeta') &= \log\left(\zeta'_0 - \gamma \frac{ k^+}{ h^+ }\right) \\
& = \log\left(\zeta'_0 \left(1 - \gamma \frac{ k^+}{ h^+ \zeta'_0}\right)\right)\\
& \approx \log(\zeta'_0)  - \gamma \frac{ k^+}{ h^+ \zeta'_0}
\end{align}
where in the last line we have used $\log(1 + x) \approx x$ for $x\ll 1$.  Plugging this back in to Eq.~\ref{eq:gamma-self-cons} and keeping only the term linear in $1/(1-N)$ gives:
\begin{equation}
\gamma \approx  \frac{1}{1-N}\left[\log(\zeta'_0) - \gamma \frac{ k^+}{h^+ \zeta'_0}\right].
\label{eq:gamma-expand-1}
\end{equation}

For $m$ defect rates, the corresponding expression is:
\begin{equation}
\gamma \approx  \frac{1}{m-N}\left[\sum_{j=1}^m\log(\zeta'_{0,j}) - \gamma \frac{ k^+}{ \sum_{j=1}^mh^+_j \zeta'_{0,j}}\right].
\label{eq:gamma-expand-mult}
\end{equation}

\subsection{Defect spacing $L_j = 0$}\label{sec:lj0}

To understand how our theory is able to correctly predict the eigenvalue of systems with adjacent defects, we write the defect transfer matrix $\MA_j$ as
\begin{equation}
\MA_j = \begin{bmatrix}
x_j & - y_j \\
1 & 0
\end{bmatrix}
\end{equation}
where 
\begin{align}
x_j &= \frac{\phi+ h_j^- + h_j^+}{h_j^+}  = \frac{ h_j^-e^{-2\pi i/N} + h_j^+e^{-2\pi i/N} + C\gamma/N}{h_j^+} \\
 y_j &= \frac{h_j^-}{h_j^+}
\end{align}
In the limit of large $N$, we have:
\begin{align}
\lim_{N\to \infty} x_j = \frac{ h_j^- + h_j^+ }{h_j^+} = 1 + \frac{h_j^-}{h_j^+} && y_j = \frac{h_j^-}{h_j^+}.
\end{align}
The eigenvalues of the product $\MA_j \MA_i$ are:
\begin{align}
\alpha^{ij}_\pm &= \frac{1}{2} \left(\pm \sqrt{({x_i} {x_j}+{y_i}+{y_j})^2-4 {y_i} {y_j}}+{x_i} {x_j}+{y_i}+{y_j}\right)
\label{eq:evpair}
\end{align}
If we can ignore the $y$ terms compared to the $x$ terms, then these reduce to:
\begin{align}
\lim_{y/x \to 0}\alpha^{ij}_1 &=0\\
\lim_{y/x \to 0}\alpha^{ij}_2 &= {x_i} {x_j}
\label{eq:evpair}
\end{align}
and the eigenvalues of $\MA_i$ become $\alpha^j_1 = x_j, \alpha^j_2 = 0$. Clearly, in this limit the eigenvalue of the product of transfer matrices is equal to the product of eigenvalues, as required, and the order of the defects will no longer matter even if they are adjacent.  The limit is fulfilled when the affinity is high not only on average but additionally along every edge in the network, so that $h^-/h^+ \sim\exp(-\aff/N)$.

Following the derivation above using $x_j$ as the eigenvalue for the defect transfer matrices, we obtain the same result as in eqs.~\ref{eq:gamma} and \ref{eq:gamma-expand-mult} with $\zeta$ replaced by $x$. Indeed, we see that in the limit that $h^-/h^+\to 0$ and $k^-/k^+\to 0$, $\zeta$ reduces to $x$. This explains how our theory can handle many adjacent defects.

\section{Constant affinity results}\label{sec:constaff}

\begin{figure}[h!]
\centering
\includegraphics[width = \linewidth]{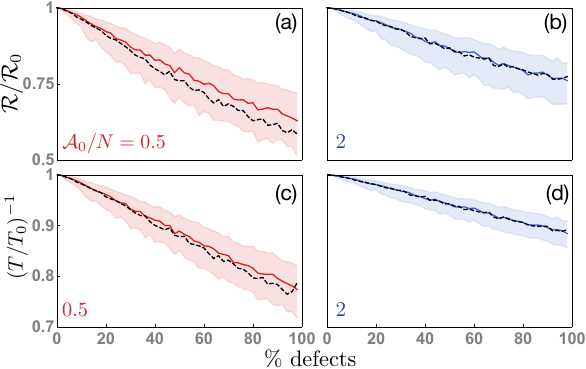}
\caption{Coherence $\R$ (a - b) and period $T$ (c - d) of an oscillator with $N = 100$ states as a function of the percent of defect rates. All counterclockwise rates are set to 1. We restrict ourselves to even numbers of defect rates. Half of the clockwise defect rates $\{h_j^+\}$ are drawn from the same distribution as in Fig.~\ref{fig:rtvsm}.  We set the other half of the defect rates to $\{\exp(\aff_0/N)^2/h_j^+\}$,  ensuring that the affinity remains constant and equal to $\aff_0$. Colors and lines have the same meaning as in Fig.~\ref{fig:rtvsm}.}
\label{fig:rtvsm-constaff}
\end{figure}

Results shown in Fig.~\ref{fig:rtvsm-constaff} confirm the bound in Ref.~\cite{Barato2017}, as the value of $\R/\R_0$ is never greater than 1, and show that $\R$ and $T$ become more robust (spread of values decreases) and predictable at high affinity.

\end{appendix}

\end{document}